\documentclass[openacc]{rstransa}%%%%where rstrans is the template name

\usepackage{dsfont}
\usepackage[numbers]{natbib}
\usepackage{txfonts}
\usepackage{footmisc}
\usepackage[percent]{overpic}
\newcommand{\ca}{\mbox{Ca\,{\textsc{ii}}~K\,}}
\newcommand{\aap}{{\it Astron. Astrophys.}}

\newcommand{\apj}{{\it Astrophys. J.}}
\newcommand{\apjs}{{\it Astrophys. J. Suppl.}}
\newcommand{\apjl}{{\it Astrophys. J. Lett.}}

\newcommand{\mnras}{{\it Mon. Not. R. Astron. Soc.}}

\newcommand{\solphys}{{\it Sol. Phys.}}

\titlehead{Research}

\begin{document}
\sloppy
 
%%%% Article title to be placed here
\title{Empirical flare energy limits for the largest historical sunspots}

\author{%%%% Author details
N.~A.~Krivova$^{1}$,
T.~Chatzistergos$^{1}$,
M.~Kazachenko$^{2,3,4}$,
and E.~I\c{s}{\i}k$^{1}$}

%%%%%%%%% Insert author address here
\address{$^{1}$Max Planck Institute for Solar System Research, Justus-von-Liebig-Weg 3,	37077 G\"{o}ttingen, Germany\\
$^{2}$ Laboratory for Atmospheric and Space Physics, University of Colorado Boulder, Boulder, CO 80303, USA\\
$^{3}$ National Solar Observatory, 3665 Discovery Drive, Boulder, CO 80303, USA\\
$^{4}$ Department of Astrophysical and Planetary Sciences, University of Colorado Boulder, 2000 Colorado Avenue, Boulder, CO 80305, USA
}

%%%% Subject entries to be placed here %%%%
\subject{astrophysics; solar system}

%%%% Keyword entries to be placed here %%%%

%%%% Insert corresponding author and its email address}
\corres{N.~A.~Krivova\\
\email{natalie@mps.mpg.de}}

%%%% Abstract text to be placed here %%%%%%%%%%%%
\begin{abstract}
Extreme solar particle events reveal that the Sun can occasionally produce eruptions significantly more energetic than those observed in the modern era.
These events are thought to originate from powerful coronal mass ejections, typically associated with large solar flares triggered by magnetic field reconnection in complex active regions.
Stellar observations indicate that Sun-like stars can host “superflares” exceeding $10^{34}$~erg roughly once per century, yet it remains uncertain whether the Sun can reach such flare energies.

We empirically estimate the upper limit of solar flare energies using statistical relations between flare-ribbon areas and released energy derived from modern observations.
By extrapolating these upper-envelope relations to the largest sunspot group recorded since 1859~--- the Great Sunspot of 8 April 1947~--- we find that exceptionally large and complex solar active regions could, in principle, produce flares with bolometric energies of a few $\times10^{34}$~erg.

\end{abstract}
%%%%%%%%%%%%%%%%%%%%%%%%%%%

%%%%%%%%%% Insert the texts which can accomdate on firstpage in the tag "fmtext" %%%%%

\begin{fmtext}

\end{fmtext}
%%%%%%%%%%%%%%% End of first page %%%%%%%%%%%%%%%%%%%%%

\maketitle

\section{Introduction}
\label{sec:intro}
Extreme Solar Particle Events (ESPEs) provide evidence that the Sun is capable of producing far more energetic eruptions than those directly observed in the modern instrumental era \cite{cliver_extreme_2022}. 
These events, identified in cosmogenic isotope records \cite[e.g.,][]{miyake_signature_2012}, are widely interpreted as originating from exceptionally strong coronal mass ejections (CMEs) and their shocks \cite{usoskin_revisited_2020,cliver_extreme_2022}. 
CMEs represent large-scale expulsions of plasma and magnetic field from the corona into interplanetary space and are the primary drivers of extreme space weather at Earth.

Solar flares are closely related to CMEs, although the relationship is not one-to-one: not all large flares produce CMEs, and not all CMEs are accompanied by major flares \citep[e.g.][]{Andrews2003,Yashiro2005}, see also Sect.~\ref{sec:results}\ref{sec:2014}.
Flares
are rapid releases of magnetic energy in the corona, powered by magnetic reconnection in solar active regions (ARs).
ARs are concentrations of strong magnetic fields associated with sunspots and plages/faculae in the chromosphere/photosphere, respectively. 
Flares emit across the entire electromagnetic spectrum, with the most energetic among them observed in white light \cite{carrington_1859,woods_solar_2004,emslie_global_2012,hayakawa_extreme_2023}. 
The total radiated flare energy is not straightforward to estimate: roughly 40–60\% emerges in the visible continuum, and 20–-30\% in the UV and EUV bands \cite{woods_contributions_2006,kretzschmar_sun_2011}.
Consequently, flare bolometric energies remain uncertain by factors of a few.
Nevertheless, the most powerful flares observed since 1975~--- such as the X25 and X43 events 
(flare classes are given on the revised NOAA scale \cite{Hudson2024})
in October–November 2003~--- reached bolometric energies of $\gtrsim 4\times10^{32}$~erg \cite{woods_contributions_2006,emslie_global_2012}.
Stellar observations suggest that the Sun might, in principle, produce much larger flares. 
Analysis of Kepler data showed that stars with near-solar parameters can host “superflares” with energies above $10^{34}$~erg at a frequency of about once per century  \cite{vasilyev_sunlike_2024}. 
Extrapolations of solar soft X-ray flare statistics suggest substantially lower occurrence rates \citep{Hudson2024} but are based on far more limited empirical coverage of extreme events; see the discussion in \cite{vasilyev_phila-2026}.

At the same time, concerns have been raised regarding whether such energetic events are physically achievable on the Sun. 
Magnetohydrodynamic (MHD) simulations scaled to the largest-ever-observed 1947 sunspot group suggest that the maximum possible solar flare energy may be limited to about 6$\times 10^{33}$~erg
\cite{aulanier_standard_2013}, while other estimates based on magnetic flux budgets of active regions and historical records similarly constrain the upper limit to below $\sim10^{34}$~erg \cite [see][for a review]{cliver_extreme_2022}.
These results imply that true ``superflares'' may be beyond solar capability.
The most powerful solar flare on record is generally considered to be the Carrington event of 1859 \cite{carrington_1859}, which produced intense aurorae and a geomagnetic storm of unprecedented strength.
Based on reconstructions from magnetic and auroral records, its flare magnitude has been estimated to lie in the range of X45--X146 \cite{cliver_extreme_2022,hayakawa_magnitude_2023,Hudson2025}, corresponding to bolometric energies of 
order a few $\times10^{32}$~erg, with some estimates approaching $\sim10^{33}$~erg,
depending on the scaling adopted to convert soft X-ray to total radiative output.
Although still below the energies of stellar superflares, the Carrington event nonetheless represents the upper bound of solar activity documented since the mid-19th century.

Understanding the physical and statistical limits of solar energy release has been an ongoing topic of interest in solar physics, see e.g., \cite{Hudson2021} for recent review and perspective on Carrington-like extreme events and their implications for space weather.
While no explicit relationship between the size of an AR and the related flare energy has been found, stronger flares originate from larger ARs, which have a higher magnetic energy budget \cite{Toriumi2019}.
Thereby the magnetic complexity of an AR is critical.

Here we revisit this question using an empirical approach.
We employ the flare-ribbon \verb+RibbonDB+ catalogue of Kazachenko et al.  \cite[][hereafter K2017]{kazachenko_database_2017}, which is based on observations from the Helioseismic and Magnetic Imager \cite[HMI;][]{scherrer_helioseismic_2012} and Atmospheric Imaging Assembly (AIA) onboard the space-based Solar Dynamics Observatory \cite[SDO;][]{pesnell_solar_2012} of more than 300 large flares between 2010 and 2016.
Flare ribbons are the bright, elongated structures observed in the chromosphere and transition region during a flare, marking the footpoints of newly reconnected magnetic field lines.
Their areas provide a direct proxy for the magnetic flux that participates in reconnection and, therefore, for the released flare energy. 
The K2017 \cite{kazachenko_database_2017} catalogue quantifies, for each event, the flare ribbon area, the associated active region area and the flare magnetic reconnection flux as a proxy for the total energy released in the flare.
From these data, the authors derived empirical scaling relations between ribbon area, peak X-ray flux and flare energy.

In this study, 
we use the same catalogue in a complementary way: we empirically characterise the upper envelope of the relation between active-region area and ribbon area (i.e., the fraction of an AR that participates in reconnection)
and then propagate those relations through the K2017 ribbon-to-energy scaling to obtain conservative, data-driven upper bounds on flare energy for the largest historical sunspot groups.
In doing so, we aim to place empirical constraints on the possible flare energies the Sun can plausibly produce. Our focus is explicitly on the extreme tail of the distribution: such events are rare, but not impossible, and they are of critical importance for assessing the likelihood and potential severity of solar superflares and ESPEs.
Strictly speaking, among the sunspots considered in our study only the 1859 and 1947 cases are historical, while the later events belong to the modern observational era.
These modern examples are included primarily as proof-of-concept benchmarks, providing consistency checks for the empirical scaling relations rather than serving as historical reconstructions.

The paper is structured as follows.
We describe the data we used and our methodology in Sect.~\ref{sec:methods}.
Our results and discussion about the flare energy limits are presented in Sect.~\ref{sec:results}.
Finally, we summarise and draw our conclusions in Sect.~\ref{sec:conclude}.

\section{Data and Method}
\label{sec:methods}

Our approach is empirical and consists of a sequence of linked steps, each building on the previous one.  
We begin by compiling records of the largest historical and modern sunspot groups, which serve as our starting point (Sect.~\ref{sec:methods}\ref{sec:hist_spots}). 
Since sunspots represent only a fraction of the magnetic flux of an active region (AR), we estimated the total AR area using two approaches: direct measurements from magnetograms and an empirical scaling relation that converts sunspot group area into total AR area (Sect.~\ref{sec:methods}\ref{sec:spot2ar}).  
Next, we use statistical relations between AR areas and flare ribbon areas, extracted from the K2017 catalogue of SDO-era flares, to estimate the possible extent of reconnection in large historical ARs (Sect.~\ref{sec:methods}\ref{sec:ribbons}).  
Finally, we employ the empirical scaling between ribbon areas and flare energies to infer the maximum plausible flare energy associated with each AR (Sect.~\ref{sec:methods}\ref{sec:energy}).  

In this way, the chain of inference connects sunspot records $\rightarrow$ AR areas $\rightarrow$ ribbon areas $\rightarrow$ flare energies.  
By applying this framework to the largest modern and historical sunspots, we aim to place empirical constraints on the upper limits of solar flare energies.

\subsection{Historical sunspot areas}
\label{sec:hist_spots}
\begin{table}[ht]
\caption{Summary of the events considered in our analysis. Flare classes are given on the revised NOAA scale \cite{Hudson2024}.}
\label{tab:events}
\centering
\begin{tabular}{llccl}
\hline
Date & Event & \multicolumn{2}{c}{Max spot area} & Flares  \\
     &       & corr, msh & proj, msd \\
\hline
 
Sep 1859 & Carrington      &  3100 & 5500 & X45--X126\\
Apr 1947 & The Great Sunspot &  6132 & 11196 &\\
Mar 1989 & Quebec			 &  4823 & 5202&  X19\\
Jul 2000 & Bastille Day   & 1429  & 2609 & X9 \\
Oct/Nov 2003 & Halloween (largest)	 &	3486 & 6039 &  X14, X25, X43 \\
& \hspace*{1.35cm} (2nd large)   &    2507 & 4423 & X4\\
Oct 2014 &AR 12192		         & 	4534 & 7683&  8$\times$X (X1--X4) + 30$\times$M-class\\%
May 2024 & Mother's Day / Gannon &  3143 & 5184&	 12$\times$X-class up to X16.5, $>$50$\times$ M\\
\hline
\end{tabular}
%\vspace*{-4pt}
\end{table}

\begin{figure}[!htb]
   \centering
    \begin{minipage}{0.98\hsize}
        \centering
        \includegraphics[width=10cm]{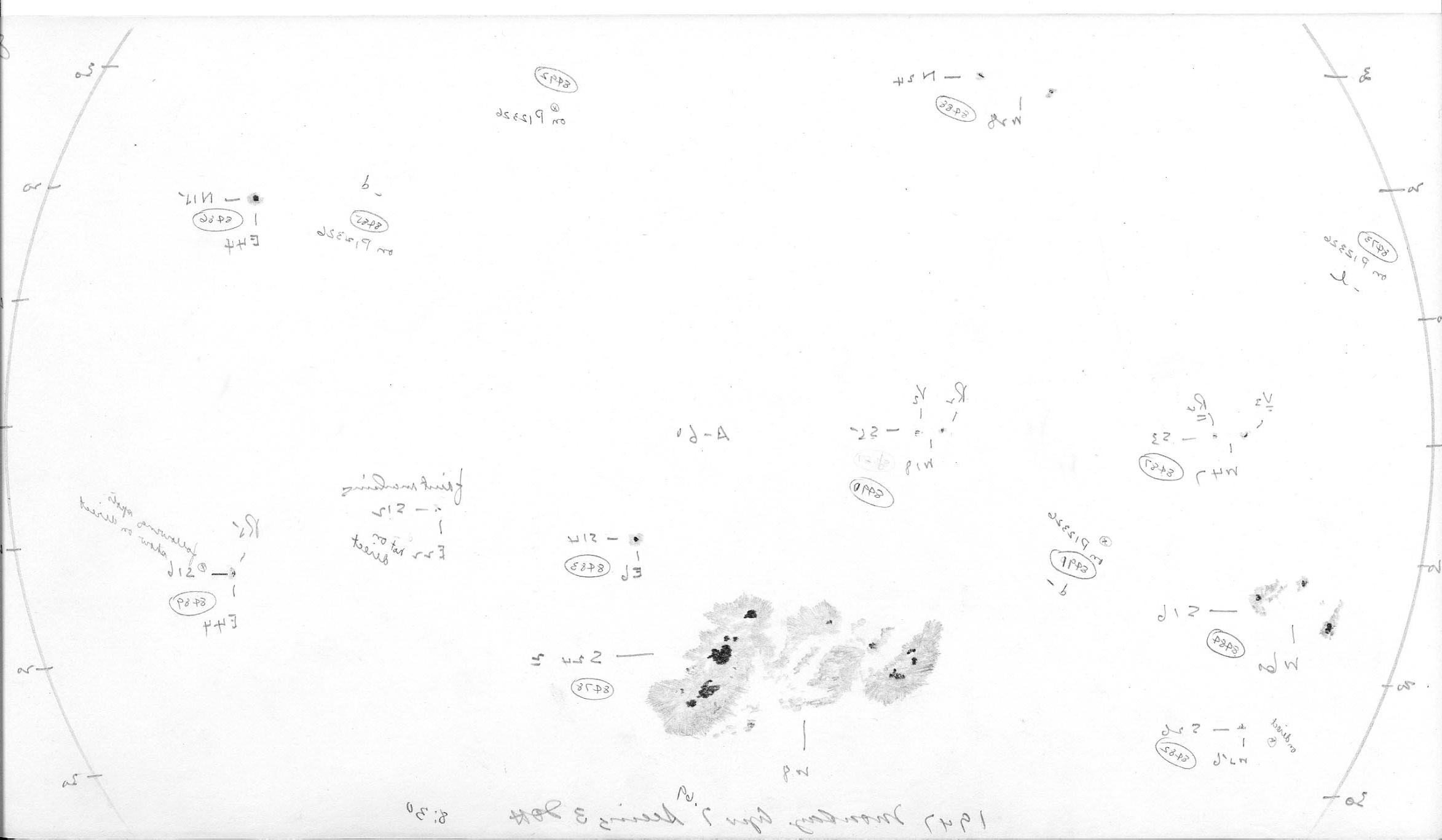}
    \end{minipage}    \begin{minipage}{0.49\hsize}
        \centering
        \includegraphics[width=7cm]{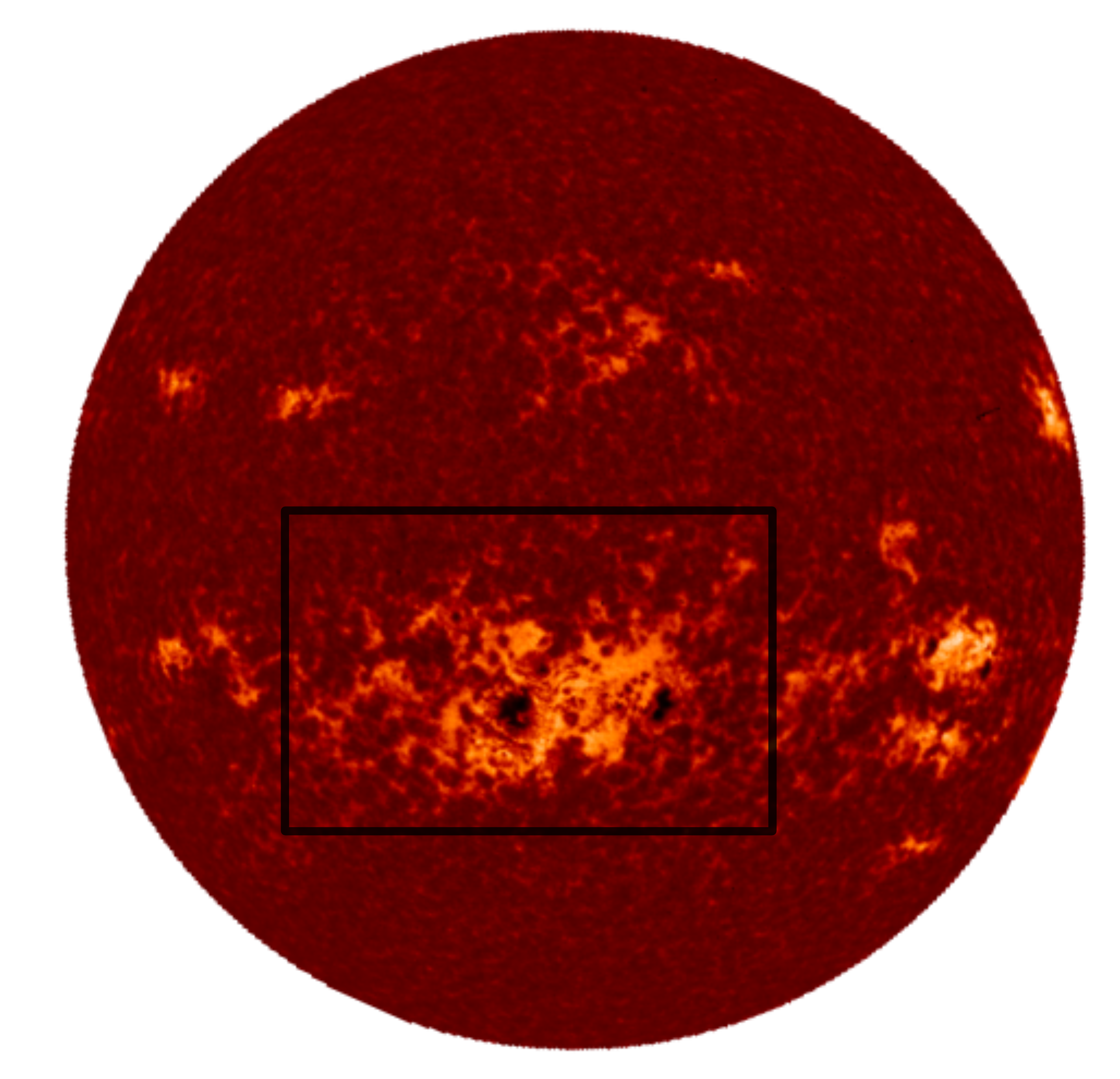}
    \end{minipage}
        \hfill
    \begin{minipage}{0.49\hsize}
        \centering
        \includegraphics[width=7cm]{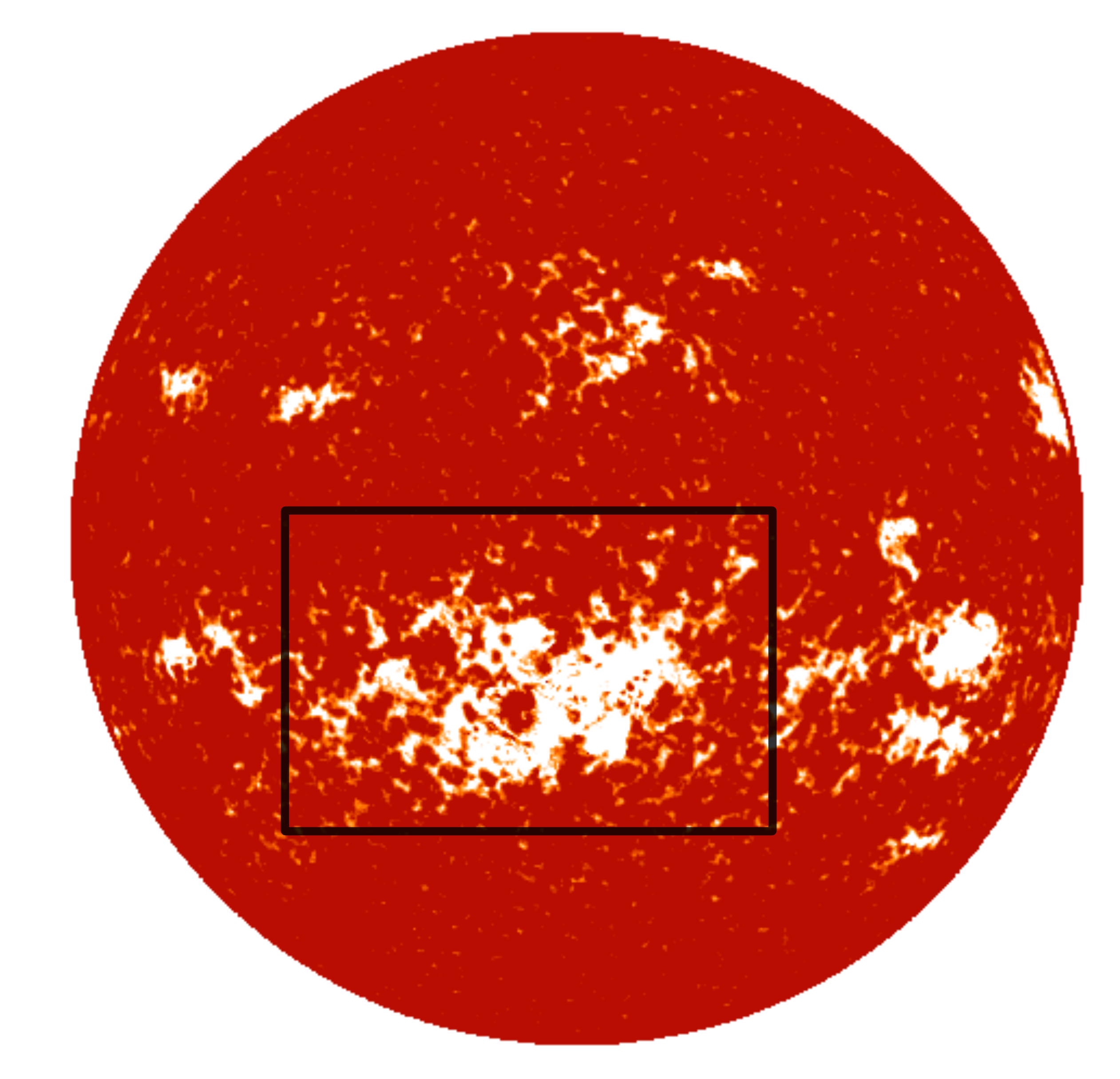}
    \end{minipage}
%\centering\includegraphics[width=2.5in]{xxxxxx.eps}
\caption{The great sunspot of 1947. Shown are Mt Wilson sunspot drawing (top), Kodaikanal \ca observation (bottom left), and reconstructed unsigned magnetogram from the Kodaikanal \ca observation (bottom right) for 7 April 1947. The \ca observation is saturated at contrast values of $\pm$1, while the magnetogram is saturated at 100 G. The black rectangle roughly marks the AR.}
\label{fig_1947}
\end{figure}

As a starting point, we selected the largest sunspot groups reported in the modern record. 
For this, we used the catalogue of Mandal et al. \cite[][hereafter M2020]{mandal_sunspot_2020}, which provides daily and group sunspot areas from 1874 onwards, as seen on the visible solar disc (projected) and corrected for the foreshortening effect, expressed in millionths of the solar disc (msd) and hemisphere (msh), respectively. 
The largest sunspot groups include, for example, the April 1947 group ($\sim$6100 msh; see Fig.~\ref{fig_1947}), the October 2003 “Halloween” group ($\sim$3500 msh; Fig.~\ref{fig_2003}), and AR 12192 of October 2014 ($\sim$4500 msh). 
These sunspot groups represent the high end of the size distribution in modern observations, but they were nevertheless associated with flaring activity well within the range of observed solar behaviour.

We note that other exceptionally large active regions have been recorded, particularly in the years immediately preceding 1947.
One example is the July 1946 sunspot group, which reached more than 4000~msh and produced an intense ground-level enhancement on 25 July 1946 \cite{forbush1946,gopalswamy_Properties_2013}.
While we do not include such cases in our quantitative analysis~--- both because the 1947 group provides a clearer reference point at the extreme upper end, and because detailed flare observations are anyway lacking~--- their occurrence illustrates that very large sunspot groups are not singular anomalies.
Rather, they appear sporadically throughout the historical record, reinforcing the plausibility of the extreme tail of the distribution.

To extend our analysis further back, we also include the Carrington sunspot group of 1859, associated with the famous geomagnetic storm. 
Direct daily sunspot area records are somewhat uncertain for this event, but several later re-analyses provide estimates. 
An approximate area of 2300 msh on 1 September 1859 is typically cited for the Carrington sunspot group \cite{newton_solar_1943}, a value also adopted by Cliver et al. \cite{cliver_extreme_2022} and earlier reviews. 
Other more recent reconstructions, based on the fraction of the solar disc covered in a drawing by R. C. Carrington for 1 September 1859, suggest a size in the range of 9--14\% of the solar diameter, corresponding to roughly 2971--3100 msh \cite{hayakawa_magnitude_2023,meadows_size_2024}. 
The sunspot group that led to the Carrington event is also included in the
drawings made by Angelo Secchi \cite{ermolli_solar_2023} between 25 August and 5 September 1859.
Based on Secchi drawings, the maximum corrected sunspot group area reached on 27 August 1859 was 3496 msh. 
We adopt 3100 msh as a conservative “upper” estimate, although it is plausible that the sunspot group had a larger area in the days preceding the Carrington event as suggested by Secchi drawings, see also Sect.~\ref{sec:results}\ref{sec:2014}. 
Based on the rough location of the sunspot group, we convert this to a projected area of $\sim$5500~msd.
In either case, the Carrington group was comparable to the largest sunspot groups of the 20th century.

We stress that our purpose is not to reconstruct the flare activity of these past events in detail.
Instead, the historical spot records provide the necessary input for a fictitious exercise: what if such large spot groups had developed according to the empirical extreme “tail” scenario that we infer from modern ribbon--AR scaling?
This allows us to assess the plausible upper bound of flare energies consistent with observed solar spot sizes.

The events explicitly included in our analysis are listed in Table~\ref{tab:events}.
For each case we report the maximum daily corrected and projected sunspot areas, expressed in millionths of the solar hemisphere (msh) and in millionths of the solar disc (msd), respectively.
We took the areas at the maximum of the group development, except for the Carrington sunspot group, for which the data are too sparse to judge when the maximum occurred.
Where available, we also note the major associated flare activity.
These cases represent the largest sunspot groups or significant flaring activity events reliably documented over the past $\sim$150 years, and they provide the empirical foundation for our extrapolation towards potential upper-limit flare energies.

%\vspace*{-5pt}

\subsection{Conversion to active region areas}
\label{sec:spot2ar}

Sunspots represent only the dark centrepieces of an AR, while the full AR encompasses a substantially larger area that includes plages and faculae.  
For our purposes, we need to estimate the total AR area, which represents the magnetic flux reservoir that can participate in flares.  
We employ two complementary approaches: (i) direct measurements from full-disc solar images and/or magnetograms, and (ii) empirical scaling of historical sunspot records using the relationship between sunspot and AR areas derived by Chatzistergos et al. \cite{chatzistergos_scrutinising_2022}.

\begin{figure}[!htb]
   \centering
    \begin{minipage}{0.49\hsize}
        \centering
        \includegraphics[width=7cm]{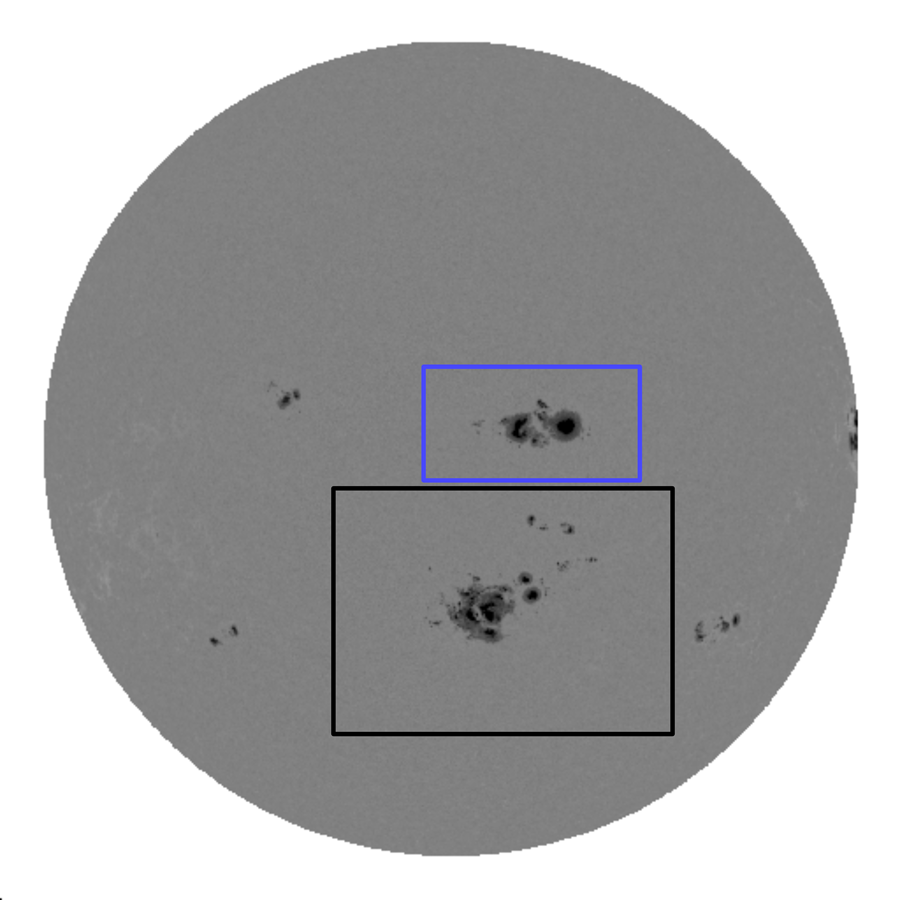}
    \end{minipage}
        \hfill
    \begin{minipage}{0.49\hsize}
        \centering
        \includegraphics[width=7cm]{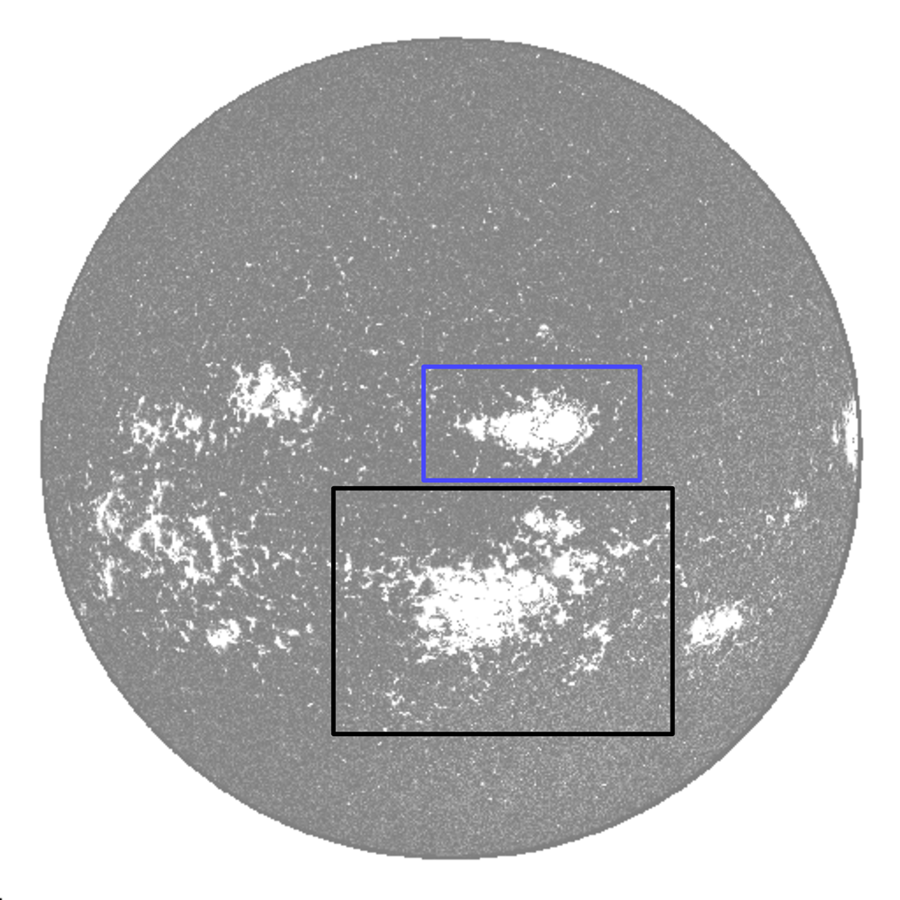}
    \end{minipage}
%\centering\includegraphics[width=2.5in]{xxxxxx.eps}
\caption{Sunspot groups on the visible solar hemisphere during the Halloween event in 2003. Shown are a continuum observation (left) and an unsigned LoS magnetogram (right) from SoHO/MDI on 29 October 2003. The magnetogram is saturated at 100 G.
The black rectangle roughly marks the larger AR, which reached a maximum spot area of $\sim 3500$~msh.
Another large AR ($\sim 2500$~msh) was present at nearly the same longitude northwards of the largest group during this period (outlined with the blue rectangle),
see Sect.~\ref{sec:results}\ref{sec:nesting} for a discussion.
}
\label{fig_2003}
\end{figure}

In the first approach, we employed full-disc line-of-sight (LoS) photospheric magnetograms and identified the target AR in each dataset (see Table~\ref{tab:full}). 
We used SDO/HMI and SoHO/MDI \citep[Michelson Doppler Imager onboard the Solar and Heliospheric Observatory;][]{Scherrer1995} LoS magnetograms as well as unsigned magnetograms reconstructed \cite{chatzistergos_recovering_2019} from Kodaikanal \cite{jha_butterfly_2024} and Meudon \cite{malherbe2023} \ca observations.  
The SDO/HMI magnetograms we used here were obtained by averaging over a 315-second period to minimise noise and p-mode–related signal fluctuations.
We approximated the radial field, $B_r$, by dividing the LoS magnetograms by the cosine of the heliocentric angle, $\mu=\cos\theta$.
To ensure consistency across instruments, all datasets were processed to an SDO/HMI-equivalent level.  
The SoHO/MDI magnetograms were cross-calibrated to SDO/HMI \cite{yeo_reconstructing_2014} using the histogram equalisation method \cite{jones_preliminary_2001}, while the \ca data were converted to unsigned magnetograms with a relationship derived for SDO/HMI magnetograms \cite{chatzistergos_recovering_2019}.
We first identified the active region of interest by selecting all interconnected pixels with $|B_r|>20$ G.
Since sunspots are not well represented in \ca data we used a white-light drawing from Mt Wilson and Meudon spectroheliograms from the wing of the \ca line (referred to as K1 at Meudon) to identify more accurately the sunspot locations and include them in the identified AR.
For consistency with the K2017 catalogue, we then selected only pixels with $|B_r|>100$ G.
We note that there might still be some small differences between our areas and those in the K2017 catalogue.
These can be due to differences in the source data and/or in the definition of the active region boundaries as well as due to intrinsic evolution of ARs on time scales of hours to days (see also Sect.~\ref{sec:results}\ref{sec:2014}). 
For example, the K2017 catalogue was based on vector magnetograms, which is not possible in our case, as such observations are not available for the largest sunspot groups included in our study.
As an illustration, Figure~\ref{fig_1947} demonstrates the exceptionally large sunspot group of April 1947
as seen in Mt.~Wilson sunspot drawing and Kodaikanal \ca observations, while Figure~\ref{fig_2003} shows the two major active regions present during the Halloween 2003 event in continuum and line-of-sight magnetograms from SoHO/MDI.

%\vspace*{-5pt}

The second approach relies on an empirical relationship between plage and sunspot areas \cite{chatzistergos_scrutinising_2022}.
This empirical relationship was determined by comparing the disc-integrated sunspot and plage projected areas by M2020 and \cite{chatzistergos_analysis_2020}, respectively resulting in a power-law relationship of the form

\begin{equation}
p_a=(-0.004\pm0.008) + (0.342\pm0.093)s_a^{(0.35\pm0.08)},
\end{equation}
where $p_a$ and $s_a$ are the plage and sunspot projected areas, respectively.
Since this relationship was derived for chromospheric plage regions, we divided them by a factor of three to estimate photospheric facular areas \cite{chatzistergos_analysis_2017,sowmya_modeling_2021}.
This factor accounts for the expansion of magnetic flux tubes with height.
The projected area of the AR, $S_{\mathrm{AR}}^{\mathrm{proj}}$, is then computed as
\begin{equation}
S_{\mathrm{AR}}^{\mathrm{proj}}=p_a/3+s_a.
\end{equation}
To correct for foreshortening, the computed areas were divided by the mean $\mu=\cos\theta$ ($\theta$ being the heliocentric angle) of the sunspot group. 

\subsection{Ribbon areas from flare statistics}
\label{sec:ribbons}

%\vspace*{-7pt}
\begin{figure}[!tbh]
   \centering
%    \begin{minipage}{0.95\hsize}
        \centering
        \includegraphics[width=0.9\hsize]{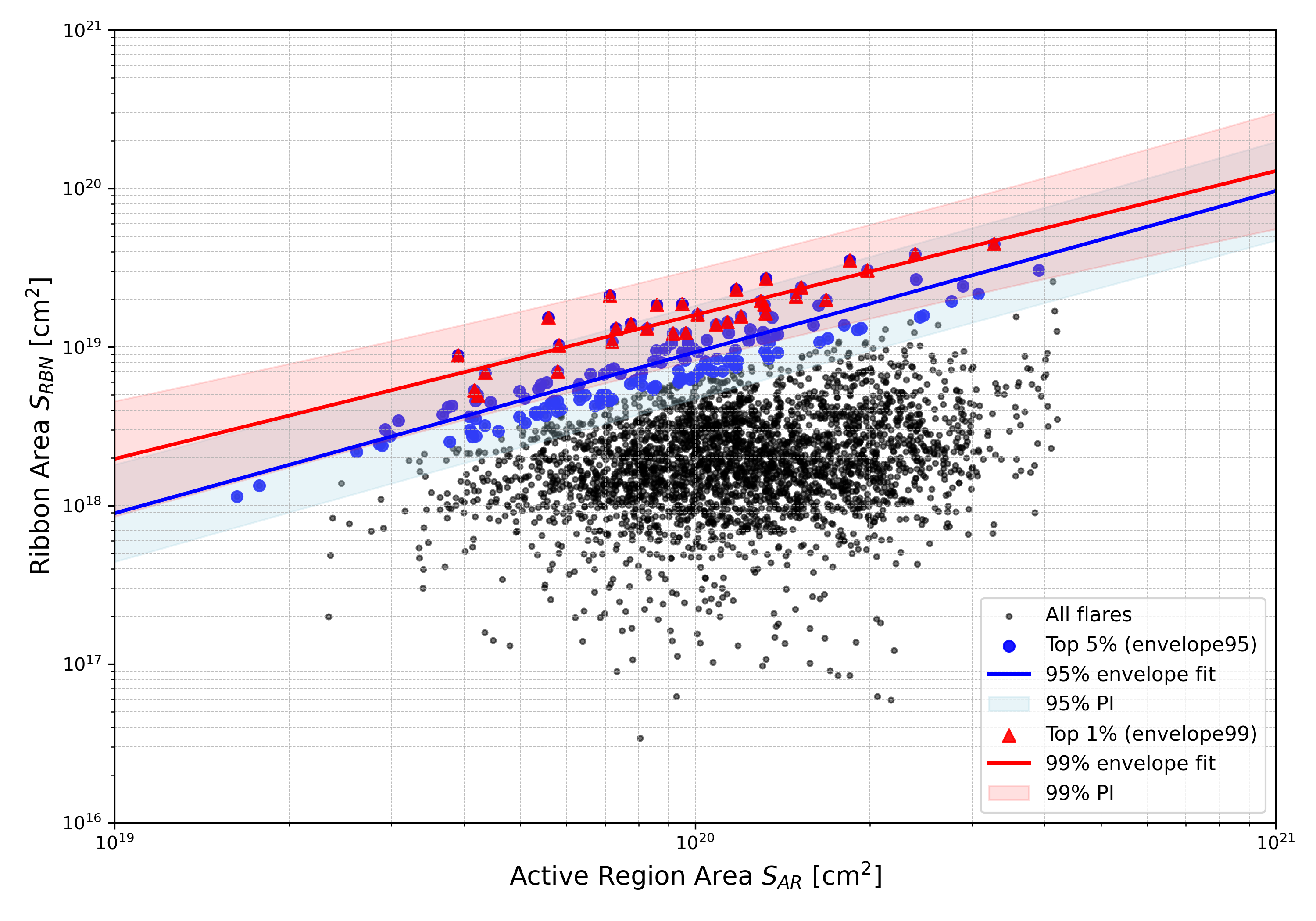}
%    \end{minipage}
%\centering\includegraphics[width=2.5in]{xxxxxx.eps}
%%% where xxxxxx name represents "figurename.eps"
\caption{
Ribbon area versus active region area for flares in the K2017 catalogue \cite{kazachenko_database_2017}.
Black points: all events; blue points: top 5\% envelope in ribbon-to-AR area ratio; red points: top 1\% envelope.
The solid blue and red lines are log–log linear fits to the 5\% and 1\% envelope subsets, respectively (that is  representing the 95th- and 99th-percentile relations.
Blue and red shaded regions: prediction intervals for the 5\% and 1\% envelope subsets, respectively. 
}
\label{fig_spot_ribbon}
\end{figure}
%\vspace*{-5pt}

For our purposes, the key aspect is the scaling between AR area and flare ribbon area. 
Flare ribbons mark the chromospheric footpoints of newly reconnected coronal loops and thus provide a direct proxy for the magnetic flux involved in a flare \cite{Forbes1984}. 
Their areas are tightly correlated with flare peak X-ray flux and thus the energy released in the event, as shown by \cite{kazachenko_database_2017,kazachenko2023}.
The K2017 catalogue, based on observations from SDO/HMI and SDO/AIA, includes measurements for 3137 flares of class C1.0 and above (of which 306 are M1.0 or greater) recorded between 2010 and 2016.
For each event, it provides the associated active-region and ribbon areas, as well as the magnetic reconnection flux.
This dataset forms the empirical basis for our analysis.

In agreement with previous studies \cite[see, e.g., the review in][]{Toriumi2019}, the distribution of ribbon areas, $S_{\mathrm{RBN}}$, relative to AR areas, $S_{\mathrm{AR}}$, exhibits large scatter: small ARs can occasionally host relatively large ribbons, and conversely large ARs can flare modestly (if at all), see Figure~\ref{fig_spot_ribbon}. 
Since our interest lies in extreme events, we focus not on the average relation but on its empirical upper envelope. 
We construct this envelope in two ways: using the 95th percentile of the $S_{\mathrm{RBN}}/S_{\mathrm{AR}}$ distribution as a function of $S_{\mathrm{AR}}$, and alternatively the 99th percentile. 
These upper-envelopes (blue and red in Figure~\ref{fig_spot_ribbon})
represent the most flare-productive regions for a given active-region size,
and serve as the basis for our subsequent estimates of maximum ribbon areas and flare energies.
The 95th-percentile envelope includes more events and therefore provides better statistics; the 99th-percentile envelope is stricter but based on fewer flares. 
In practice, the 95th-percentile case also serves as a consistency check that the 99th-percentile result is not unduly affected by small-number statistics.
Indeed, the slope of the 95th-percentile fit is slightly steeper (although well within the fit uncertainties) than that of the 99th-percentile fit (see Figure~\ref{fig_spot_ribbon}).
This could reflect a physical effect, such as a saturation of ribbon sizes in the largest ARs, but it may also be a statistical bias due to the relative scarcity of huge ARs and extreme events. 
Thus, we have two complementary estimates:  
{\em(i)} a ``95-envelope'' scaling, which captures the bulk of large flares, and  
{\em(ii)} a rarer ``99-envelope'' scaling, which emphasizes the extreme tail.

We fit both percentiles with power-law relations of the form
\begin{equation}
\label{eq:Srbn}
S_{\mathrm{RBN}} = a \, S_{\mathrm{AR}}^{\,b},
\end{equation}
where $S_{\mathrm{RBN}}$ and $S_{\mathrm{AR}}$ are the ribbon and AR areas, respectively.
In addition, we compute the corresponding prediction intervals (PI) of the fits, both for 95\% and 99\% envelopes, which quantify the statistical uncertainty of the extrapolation.
Thus, for each AR we obtain not only a “best” estimate from the 95- and 99-percentile envelopes, but also conservative upper bounds from the respective prediction intervals.

\subsection{Flare energy scaling}
\label{sec:energy}

To translate flare ribbon areas into flare energies, we rely on the empirical relations established by K2017 \cite{kazachenko_database_2017}.  
Their catalogue quantifies ribbon areas and reconnection fluxes for 3137 flares of GOES class C1.0 and greater (including $306$ M- and X- class flares) observed with SDO/HMI and SDO/AIA.  
By combining ribbon measurements with the underlying magnetic field, they derived the reconnected magnetic flux and showed that it correlates tightly with GOES soft X-ray peak flux, following approximately  
\begin{equation}
I_{\mathrm{X,peak}} \propto \Phi_{\mathrm{ribbon}}^{1.5},
\end{equation}  
where $I_{\mathrm{X,peak}}$ is the GOES peak flux and $\Phi_{\mathrm{ribbon}}$ the reconnection flux or the total unsigned magnetic flux swept by the flare ribbons.  
This provides an observationally grounded link between flare ribbon properties and flare radiative output.
The total magnetic energy release can then be approximated as
\begin{equation}
\label{eq:energy}
    E \approx S_{\mathrm{RBN}}^{1.5} * f * B^2 / (8\pi * 2),
\end{equation} 
where $B$ is the mean magnetic field strength in the ribbon region, and $f$ describes the fraction of magnetic energy that actually participates in reconnection and the fraction of the released energy that appears as flare radiation.

In K2017 \cite{kazachenko_database_2017}, the mean ribbon magnetic field was found to be typically in the range 200--1300~G (see their Figure~10). 
Following K2017, we adopt $B = 1000$~G for consistency. 
Because the flare energy scales as $E\propto B^2$, variations in the ribbon magnetic field directly translate into substantial variations in inferred flare energy.
The strongest flares are naturally expected to arise from active regions with ribbon magnetic fields near the upper end of the observed distribution. 
Therefore, the adopted value 
$B=1000$\,G represents a plausible high-field case rather than an extreme choice.
The factor $f$ depends on the fraction of the free magnetic energy that participates in the flare and then gets converted into radiative + non-radiative + CME kinetic components.
K2017 \cite{kazachenko_database_2017} used $f = 1.0$.
According to \cite{emslie_global_2012,aschwanden_energetics_2017}, up to 10--30\% of the released magnetic energy can appear in radiative form, while the rest contributes to particle acceleration and CME dynamics.
Interested in the potential higher limits, we adapt $f \approx 0.3$.

\begin{figure}[!tbh]
   \centering
%    \begin{minipage}{0.95\hsize}
        \centering
        \includegraphics[width=0.9\hsize]{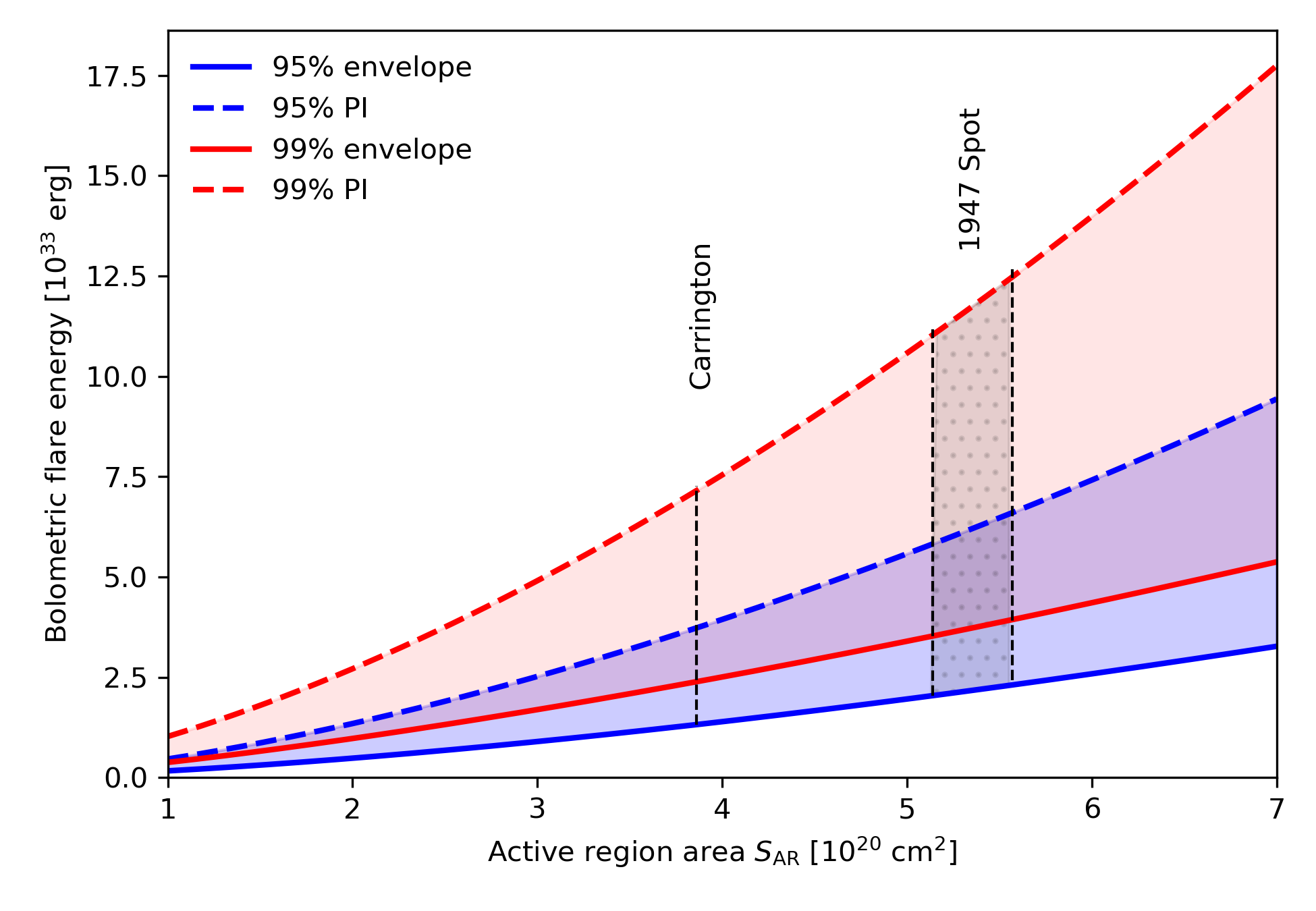}
%    \end{minipage}
\caption{Bolometric flare energy as a function of active-region area.
Solid curves show the empirically derived 95\% (blue) and 99\% (red) envelope relations between AR area and bolometric flare energy, obtained from Eqs.~\ref{eq:Srbn} and \ref{eq:energy}. 
Dashed curves indicate the corresponding upper prediction intervals for each envelope.
Shaded regions highlight the range between the envelope and its prediction interval.
Vertical dotted lines mark AR areas of the Carrington (1859) and 1947 event.
For the 1947 event, both image-based and empirically inferred estimates are shown, with the corresponding range indicated by hatching.
}
\label{fig_AR_energy}
\end{figure}
%\vspace*{-5pt}

In practice, we use the ribbon-area–energy scaling reported by K2017 as our baseline.  
To focus on the most energetic events, we use the ribbon area estimates for the upper envelopes of the distribution (95th and 99th percentiles).
Extrapolating these empirical upper relations to the largest historical active regions allows us to estimate the maximum flare energies that the Sun could plausibly produce.  

We emphasise that this extrapolation carries important caveats.  
First, the envelope slopes differ slightly between the 95- and 99-percentile cases; this may reflect a physical saturation effect at very large ARs, or it may simply arise from limited statistics at the extreme tail of the distribution.  
Second, the relation between ribbon areas and flare energies is itself subject to scatter and systematic uncertainty, since it is based on a limited number of modern events.  
Therefore, our approach does not yield precise predictions for individual historical flares, but instead provides plausible upper bounds on the flare energies that could in principle be generated by the largest observed active regions.

The percentile envelopes used here therefore have a statistical rather than deterministic meaning.
The 95th- and 99th-percentile envelopes represent the upper tails of the empirically observed ribbon--AR area relationship and characterise rare, but still observed cases, in which magnetic flux is converted into flare-participating ribbons with higher-than-usual efficiency.
They should not be interpreted as strict physical limits for individual active regions.
Rather, they provide probabilistic upper bounds that reflect the combined effects of magnetic complexity, reconnection geometry, and variability in flare efficiency.
The associated prediction intervals quantify the intrinsic scatter of this relationship and illustrate the range of flare energies that may arise for a given AR size under particularly favourable conditions.

To visualise how the inferred flare-energy bounds scale with AR size, Fig.~\ref{fig_AR_energy} shows the peak bolometric flare energy as a function of AR area based directly on the fitted scaling relations (Eqs.~\ref{eq:Srbn} and \ref{eq:energy}).
We plot both the 95th- and 99th-percentile envelope relations together with their corresponding prediction intervals.

\section{Results and discussion}
\label{sec:results}

\begin{table}[ht]
\caption{Summary of the active regions considered in this study and derived potential properties of flares associated with them.
Columns list the year of each event (we considered two largest ARs associated with the Halloween event in 2003), the reported maximum sunspot area, and the corresponding quantities obtained from image-based and empirically derived scaling relations.
For the image-based estimates, $S_{\mathrm{AR}}^{\mathrm{img}}$ denotes the total active-region area measured from historical or modern observations (sources indicated below the header), while $S_{\mathrm{RBN}}^{\mathrm{img}}$ and $E^{\mathrm{img}}$ represent the corresponding flare-ribbon areas and bolometric flare energies estimated from the scaling relations, shown for the 95th- and 99th-percentile envelopes and their prediction intervals (PI). 
For the empirically derived quantities, $S_{\mathrm{AR}}^{\mathrm{emp}}$~--- and subsequently $S_{\mathrm{RBN}}^{\mathrm{emp}}$ and $E^{\mathrm{emp}}$~--- are obtained from the empirical spot--AR relation \cite{chatzistergos_scrutinising_2022}. 
All areas are expressed in units of $10^{20}$~cm$^2$ and energies in $10^{33}$~erg.
}
\label{tab:full}
\centering
%\scriptsize
\setlength{\tabcolsep}{3pt}
\renewcommand{\arraystretch}{1.0}
\begin{tabular}{lcccccccc}
\hline
Spot Year & 1859 & 1947 & 1989 & 2000 & 2003-1 & 2003-2 & 2014 & 2024 \\
\hline\hline\\
Spot area (msh) & 3100.00 & 6132.00 & 4822.58 & 1428.84 & 3486.34 & 2506.84 & 4533.62 & 3142.86 \\
\hline\\[-5pt]
{\bf\em AR area derived from images}\\
&--&Koda&Meudon&MDI&MDI&MDI&HMI&HMI\\
$S_{\mathrm{AR}}^{\mathrm{img}}$ [$10^{20}$ cm$^2$] & -- & 5.14 & 4.35 & 1.57 & 2.82 & 1.23 & 3.29 & 2.09 \\
$S_{\mathrm{RBN}}^{\mathrm{img}}$ (95\% best) [$10^{20}$ cm$^2$] & -- & 0.49 & 0.41 & 0.15 & 0.27 & 0.11 & 0.31 & 0.20 \\
$S_{\mathrm{RBN}}^{\mathrm{img}}$ (95\% PI) [$10^{20}$ cm$^2$] & -- & 0.98 & 0.83 & 0.29 & 0.53 & 0.23 & 0.62 & 0.39 \\
$S_{\mathrm{RBN}}^{\mathrm{img}}$ (99\% best) [$10^{20}$ cm$^2$] & -- & 0.70 & 0.61 & 0.24 & 0.41 & 0.19 & 0.47 & 0.31 \\
$S_{\mathrm{RBN}}^{\mathrm{img}}$ (99\% PI) [$10^{20}$ cm$^2$] & -- & 1.51 & 1.27 & 0.47 & 0.83 & 0.37 & 0.96 & 0.62 \\
% ENERGIES
$E^{\mathrm{img}}$ (95\% best) [$10^{33}$ erg] & -- & 2.04 & 1.58 & 0.33 & 0.82 & 0.23 & 1.04 & 0.52 \\
$E^{\mathrm{img}}$ (95\% PI) [$10^{33}$ erg] & -- & 5.82 & 4.49 & 0.93 & 2.29 & 0.64 & 2.91 & 1.45 \\
$E^{\mathrm{img}}$ (99\% best) [$10^{33}$ erg] & -- & 3.53 & 2.81 & 0.70 & 1.56 & 0.51 & 1.92 & 1.04 \\
$E^{\mathrm{img}}$ (99\% PI) [$10^{33}$ erg] & -- & 11.03 & 8.56 & 1.91 & 4.48 & 1.37 & 5.63 & 2.90 \\
\hline\\[-5pt]
%EMPIRICAL
{\bf\em AR area derived empirically}\\[5pt]
$S_{\mathrm{AR}}^{\mathrm{emp}}$ [$10^{20}$ cm$^2$] 
& 3.86 & 5.57 & 6.17 & 2.57 & 4.16 & 3.47 & 4.84 & 4.03 \\
%$S_{\mathrm{AR}}^{\mathrm{emp, upper}}$ [$10^{20}$ cm$^2$]
%& 7.21 & 9.19 & 11.62 & 5.45 & 7.63 & 6.71 & 8.53 & 7.59 \\
$S_{\mathrm{RBN}}^{\mathrm{emp}}$ (95\% best) [$10^{20}$ cm$^2$] 
& 0.37 & 0.53 & 0.59 & 0.24 & 0.39 & 0.33 & 0.46 & 0.38 \\
$S_{\mathrm{RBN}}^{\mathrm{emp}}$ (95\% PI) [$10^{20}$ cm$^2$]
& 0.73 & 1.07 & 1.19 & 0.48 & 0.79 & 0.65 & 0.93 & 0.76 \\
%& 1.40 & 1.80 & 2.31 & 1.05 & 1.49 & 1.30 & 1.67 & 1.48 \\
$S_{\mathrm{RBN}}^{\mathrm{emp}}$ (99\% best) [$10^{20}$ cm$^2$] 
& 0.54 & 0.76 & 0.83 & 0.38 & 0.58 & 0.49 & 0.67 & 0.56 \\
$S_{\mathrm{RBN}}^{\mathrm{emp}}$ (99\% PI) [$10^{20}$ cm$^2$]
& 1.13 & 1.63 & 1.81 & 0.75 & 1.22 & 1.01 & 1.42 & 1.18 \\
% & 2.13 & 2.74 & 3.51 & 1.60 & 2.26 & 1.98 & 2.54 & 2.25 \\
% ENERGIES
$E^{\mathrm{emp}}$ (95\% best) [$10^{33}$ erg]
& 1.32 & 2.31 & 2.70 & 0.71 & 1.48 & 1.12 & 1.87 & 1.41 \\
$E^{\mathrm{emp}}$ (95\% PI) [$10^{33}$ erg]
& 3.73 & 6.59 & 7.74 & 1.98 & 4.19 & 3.15 & 5.31 & 3.98 \\
%& 9.88 & 14.46 & 20.90 & 6.38 & 10.81 & 8.84 & 12.87 & 10.71 \\
$E^{\mathrm{emp}}$ (99\% best) [$10^{33}$ erg]
& 2.39 & 3.93 & 4.52 & 1.37 & 2.64 & 2.06 & 3.26 & 2.53 \\
$E^{\mathrm{emp}}$ (99\% PI) [$10^{33}$ erg]
& 7.15 & 12.46 & 14.59 & 3.90 & 8.00 & 6.08 & 10.08 & 7.62 \\
%& 18.56 & 27.12 & 39.20 & 12.07 & 20.30 & 16.64 & 24.14 & 20.12 \\
\hline
\end{tabular}
\end{table}

\subsection{Overview of analysed events}
\label{sec:overview}

Table~\ref{tab:events} summarises the eight representative large sunspot groups considered in this study, ranging from the Carrington event of 1859 to the exceptionally large AR of 1947 and more recent strong solar eruptive storms such as the Halloween 2003 and October 2014 events.  
For each event, we give dates, the observed sunspot areas (projected on the disc and corrected for the foreshortening) and the strongest flares associated with them.

Table~\ref{tab:full} lists the analysed events and the corresponding derived quantities.
For each, we estimated the active-region and ribbon areas using two approaches: (1) areas derived directly from  images (“img”) and (2) areas obtained from the empirical spot-to-AR conversion (“emp”) described in Sect.~\ref{sec:methods}\ref{sec:spot2ar}. 
The corresponding flare energies were then computed using the flare energy scaling (Sect.~\ref{sec:methods}\ref{sec:energy}) applied on ribbon areas derived from the upper-envelope relations between AR and ribbon areas (Sect.~\ref{sec:methods}\ref{sec:ribbons}).
The resulting energies range from a few $10^{32}$~erg for smaller of the analysed active regions to nearly $10^{34}$~erg for the largest historical spots.
Throughout this section, we compare observed events primarily with the ``best'' (95th-envelope) scaling, which represents the typical scaling between AR and ribbon areas for the top 5\% strongest events in the K2017 catalogue; the prediction intervals for the 95\% and 99\% envelopes are treated as statistical extrapolations beyond the directly observed regime aiming to assess the potential limits of the rare extreme events.

\subsection{Comparison with AR 12192 (October 2014)}
\label{sec:2014}
AR 12192, which was the source of multiple strong flares extensively studied in the literature \cite[e.g.,][]{sun_2015,Sarkar_2018,thalmann2019}, is the only one from our sample  which is also covered by the K2017 catalogue.
We therefore start our discussion from this event, which allows a direct comparison of our predictions with observations.

\begin{table*}[ht]
\centering
\caption{Comparison of active-region and ribbon areas for AR\,12192 (Oct 2014).
Values in units of $10^{20}$\,cm$^2$. ``Kaz'' = K2017 \cite{kazachenko_database_2017} measured values;
``img'' = our image-based AR area; ``emp'' = our empirical spot$\to$AR conversion.}
\label{tab:AR12192_comp}
\begin{tabular}{lcccc}
\hline
Date & 24.10  & 25.10  & 26.10  & 27.10  \\
\hline %\hline\\[-2mm]
\textbf{Spot areas} \\
Spot area, corr. (msh) & 
4373.74 & 4533.62 & 3878.84 & 3573.08 \\
Spot area, proj. (msd) & 
8384.40 & 7683.39 & 6065.94 & 4654.26 \\[5pt]
\textbf{Active-region areas} \\
$S_{AR}^{\rm Kaz}$ & 3.27 & 3.91 & 4.10 & 4.20 \\[3pt]
$S_{AR}^{\rm img}$ & 3.11 & 3.29 & 3.52 & 3.83 \\
$S_{AR}^{\rm emp}$ & 4.50 & 4.84 & 4.62 & 4.82 \\[5pt]
\textbf{Ribbon areas} \\
$S_{RBN}^{\rm Kaz}$ & 0.45 & 0.30 & 0.26 & 0.17 \\[3pt]
$S_{RBN}^{\rm img}$ (95\% best) & 0.29 & 0.31 & 0.33 & 0.36 \\
\quad (95\% PI) &  0.58 & 0.62 & 0.66 & 0.73 \\
\quad (99\% best) & 0.45 & 0.47 & 0.50 & 0.54 \\
\quad (99\% PI) & 0.91 & 0.96 & 1.03 & 1.12 \\[3pt]
$S_{RBN}^{\rm emp}$ (95\% best) & 
0.43 & 0.46 & 0.44 & 0.46 \\
\quad (95\% PI) & 0.86 & 0.93 & 0.88 & 0.92 \\
\quad (99\% best) & 0.62 & 0.67 & 0.64 & 0.66 \\
\quad (99\% PI) & 1.31 & 1.42 & 1.35 & 1.41 \\
\hline
\end{tabular}
\end{table*}

Table~\ref{tab:AR12192_comp} summarises the active-region and ribbon areas for four days between 24 and 27~October, when AR\,12192 reached its maximum development.
The sunspot group reached its maximum corrected area on 25~October \cite{mandal_sunspot_2020}, which is therefore the value used in our summary Table~\ref{tab:full}, whereas the K2017 active-region areas peaked one to two days later.
This offset is consistent with the expected evolution of a large emerging region, where the magnetic footpoints continue to expand even when the sunspots begin to decay. 
Notably, the flare ribbons reached their largest extents slightly earlier, around 24--25~October, possibly reflecting enhanced magnetic complexity during the phase of strongest flux emergence.
We include these four days to capture this evolution and to provide a more comprehensive comparison between measured and predicted quantities.
Obviously, such a complex temporal evolution is not reflected in our simplified analysis, which is based on the day when the sunspot area reached its maximum. 
As a result, the actual ribbon areas may in some cases have been larger at slightly different stages of the region development.

The active-region areas given by K2017 range between $3.3$ and $4.2\times10^{20}$~cm$^2$, in good agreement with both our image-based estimates ($3.1$–$3.8\times10^{20}$~cm$^2$) and our empirical conversion from sunspot to AR areas ($4.5$–$4.8\times10^{20}$~cm$^2$).
The image-based areas are slightly lower than the measured ones. 
We think that this is due to the different approach used by K2017 to derive the areas.
Whereas they considered a box of fixed size encompassing the AR but potentially also including some other magnetic regions, we identified the boundaries of each given AR individually.
Our empirical estimates, in contrast, tend to lie at the higher tail of the measurements.
In general, our two approaches encompass fairly well the areas in the catalogue, and
the consistency across all three methods indicates that our AR–spot calibration is well aligned with K2017.

For the flare ribbons, K2017 report areas between $0.17$ and $0.45\times10^{20}$~cm$^2$.
Our image-based upper-envelope predictions yield ``expected'' ribbon areas of $0.29$–$0.36\times10^{20}$~cm$^2$ for the 95\% case and $0.45$–$0.54\times10^{20}$~cm$^2$ for the 99\% case, closely matching the observed range.
The corresponding prediction intervals (95\%: $0.58$–$0.73\times10^{20}$~cm$^2$; 99\%: $0.91$–$1.12\times10^{20}$~cm$^2$) extend modestly above the measured values, as expected 
since these upper relations are constructed by statistically propagating the intrinsic scatter and prediction intervals.
Our empirically derived ribbon areas show similar behaviour, with ``best'' 95\% envelope predictions of $0.43$–$0.46\times10^{20}$~cm$^2$ and 99\% envelope values of $0.62$–$0.66\times10^{20}$~cm$^2$, while
the prediction intervals (95\%: $0.86$–$0.92\times10^{20}$~cm$^2$; 99\%: $1.31$–$1.41\times10^{20}$~cm$^2$) are slightly higher than for the image-based estimates. 

The overall agreement between the observed ribbon areas and both our image-based and empirical ``best'' estimates provides confidence that the method captures the typical AR–ribbon scaling correctly.
Since the upper estimates are constructed by systematically propagating the scatter and prediction intervals, the fact that the baseline predictions already agree with the data suggests that our extrapolated estimates for the rarest, tail-end events are also realistic. 
In this sense, the consistency with AR 12192 lends support to the use of our upper-envelope predictions in assessing the extreme limits of solar flare energies.

We note that AR 12192 was characterised by an unusually high degree of flare confinement \cite{Chen2015,Sun2015,Thalmann2015},
with all of its X-class flares (X1--X4.4) confined and only weak associated SEP activity.
This demonstrates that large sunspot area and strong magnetic fields do not automatically produce the most extreme eruptive events.
Strong overlying magnetic fields can suppress CME formation, thereby limiting efficient particle escape \citep{Yashiro2005,Sun2015,Thalmann2015,AlvaradoGomez2018,Li2020}.
A higher degree of confinement in very large and magnetically complex active regions could therefore reduce the probability that the highest-energy flares are accompanied by fast CMEs and extreme SEP events, potentially contributing to the rarity of ESPEs.
Nevertheless, large sunspot area does not necessarily imply confinement.
The larger 1989 Quebec region (see Table~\ref{tab:events}) produced an eruptive X19 flare with substantial space-weather impact, indicating that magnetic topology and coronal field structure play a decisive role in determining whether stored magnetic energy is released eruptively or remains confined.
Our scaling relations therefore constrain the magnetic energy potentially available in large active regions, while the realised flare magnitude, CME occurrence, and SEP production depend on the magnetic configuration and eruption dynamics.

\subsection{The Halloween 2003 event}
\label{sec:2003}

A particularly important test case is the Halloween storm of late October and early November 2003 (see Figure~\ref{fig_2003}), during which multiple extremely strong flares were observed from AR\,10486.
This includes the X25 and X43 flares. 
Unlike most historical flares, the bolometric energy of 
several major flares during the Halloween 2003 period was directly constrained using space-based total solar irradiance (TSI) observations \cite{woods_solar_2004}.
After applying limb-darkening corrections to these measurements \cite{Emslie2012}, the 4~November 2003 event~-- reclassified as an X43 flare on the revised NOAA scale \cite{Hudson2024}~-- had the largest inferred bolometric energy of the Halloween events, approximately $4.3\times10^{32}$~erg.
Two other major flares had inferred bolometric energies of $\sim3.6\times10^{32}$~erg (28 October 2003) and $\sim1.4\times10^{32}$~erg (29 October 2003), respectively \cite{Emslie2012}.
The associated uncertainties are, however, large, reaching up to $\sim$90\%.

Our method predicts flare energies for AR\,10486 of $E^{\mathrm{img}} \approx 8\times10^{32}$~erg for the ``best'' 95th-envelope scaling, of the same order as the directly measured bolometric energy of the most energetic Halloween flare on 4 October ($\sim4.3\times10^{32}$~erg, 65\% uncertainty).
The estimated range for the less likely but stronger events is up to $2.3\times10^{33}$~erg (95\%~PI) and $4.5\times10^{33}$~erg (99\% PI) using the image-based areas (Table~\ref{tab:full}). 
Using the empirical areas, the corresponding predictions are slightly higher: $1.5\times10^{33}$~erg (best 95th envelope), $4.2\times10^{33}$~erg for 95\%~PI, and up to $8\times10^{33}$~erg for the 99\% prediction interval.  

During the Halloween storm, the dominant AR\,10486 produced the most powerful events (X25, X43), but another unusually large group (AR\,10488) was also present on the disc, with an area exceeding 2500~msh (see Fig.~\ref{fig_2003}). 
We include both in our analysis (Tables~\ref{tab:events},~\ref{tab:full}), since they jointly contributed to the remarkable activity of late October and early November 2003. 
We return to these two regions in Sect.~\ref{sec:results}\ref{sec:nesting}.

Taken together, the 2003 events provide a rare case where direct bolometric flare energies can be compared with empirical scaling predictions, offering strong empirical support for our framework and for its extrapolation to the most extreme solar active regions.

\subsection{The Bastille Day 2000 event}
\label{sec:bastille}

Another noteworthy case is the so-called Bastille Day flare of 14~July~2000, originating from AR\,9077.  
This active region is the smallest included in our sample, with a maximum sunspot-group area of only about 1400~msh (Table~\ref{tab:events}).  
Despite its modest size compared with the giant spots of 1947 or 2003, AR\,9077 produced one of the most prominent space-weather events of the modern era, including a fast halo CME and a strong ground-level enhancement \cite{shea_smart_2012,cliver2013}.

Kazachenko et~al.~\cite{Kazachenko2012} analysed multi-wavelength data and estimated the bolometric energy of the flare to be $(3.4\pm1.7)\times10^{32}$~erg.  
From Table~\ref{tab:full}, our empirical scaling for AR\,9077 predicts a flare energy of $7.1\times10^{32}$~erg under the ``best'' 95th-envelope relation, $1.98\times10^{33}$~erg and $3.90\times10^{33}$~erg within the 95\% and 99\% prediction intervals, respectively.
The values obtained from direct image measurements are $3.3\times10^{32}$~erg for the ``best'' 95th-envelope relation~--- in perfect agreement with the observed value, and $9.3\times10^{32}$~erg and $1.91\times10^{33}$~erg within the 95\% and 99\% prediction intervals, respectively.
The observed bolometric energy therefore falls well within our predicted distribution, confirming that even relatively small active regions can produce flares of a few $10^{32}$~erg.

This example highlights two key points.  
First, comparatively modest active regions can still yield large events, underscoring the wide intrinsic variability of solar-flare productivity.  
Second, the consistency of our statistical scaling with the observed bolometric energy lends further support to the validity of our approach across a broad range of AR sizes. 

\subsection{The Great Sunspot of 8~April~1947}
\label{sec:1947}

The sunspot group of April~1947 (AR\,14886) remains the largest on record, with an estimated maximum corrected area of about 6100~msh \cite{mandal_sunspot_2020}.  
Figure~\ref{fig_1947} shows solar observations from 7~April~1947 highlighting this huge region in white light and in \ca.

Our empirical scaling implies a typical flare energy of $2.31\times10^{33}$~erg under the ``best'' 95th-envelope relation, and $6.59\times10^{33}$~erg to $1.25\times10^{34}$~erg within the 95\% and 99\% prediction intervals, respectively.
The values obtained from direct images are only very slightly lower.
Thus, a flare from a region of this size could plausibly reach energies of over $10^{34}$~erg if it represented an extreme outlier at the upper end of the empirical distribution.  
This range overlaps with the lower limit of the ``superflare'' regime inferred from stellar statistics \cite{shibayama_superflares_2013,vasilyev_sunlike_2024}, suggesting that the Sun could, in principle, reach this level~--- albeit extremely rarely.
No bolometric measurements or flare reports exist for 1947.
Whether such an event actually occurred is unknown, but the energy potential was clearly present.

\subsection{The Carrington event of 1859}
\label{sec:carrington}

The 1859 ``Carrington event'' is considered the benchmark for extreme solar activity in the historical record. 
Based on archival sunspot drawings, the associated group reached a maximum corrected area of about 3100--3500~msh \cite[][see Sect.~\ref{sec:methods}\ref{sec:hist_spots}]{hayakawa_magnitude_2023,meadows_size_2024,ermolli_solar_2023}, placing it among the largest active regions observed since the mid-19th century.
We used the more conservative value of 3100~msh.
Applying our empirical scaling yields a typical flare energy 
of $\sim 1.3\times10^{33}$~erg under the 95th-envelope relation (Table~\ref{tab:full}). 
This value is fully consistent with independent estimates for the Carrington flare X45--X146 \cite{cliver_extreme_2022,hayakawa_magnitude_2023,Hudson2025}, corresponding to bolometric energies approaching $\sim10^{33}$~erg, although subject to substantial uncertainties in converting soft X-ray class to total radiative output.
The 99\%-envelope prediction interval indicates that, under exceptionally favourable conditions, the flare energy could plausibly reach up to $\sim7\times10^{33}$~erg.

Although no direct bolometric measurement is available, the agreement between our extrapolated energy and these historical inferences supports the physical plausibility of the Carrington event representing one of the most energetic flares in recorded solar history.  
It also provides a useful empirical anchor for assessing the upper limit of solar flare energy.

\subsection{Implications for extreme solar events}
\label{sec:implications}

Taken together, the results from Table~\ref{tab:full} provide a consistent empirical picture of the upper limits of solar flare productivity.  
For the modern, well-observed events (2000--2014), flare energies derived from the image-based scaling relations agree within a factor of two with the empirically derived values, lending confidence to the overall consistency of the approach.  
The difference between the two methods decreases systematically with increasing active-region size, suggesting that our empirical relations possibly capture the large-scale magnetic and geometric properties of major regions particularly well.  
The ``best'' (95th-envelope) relations reproduce the typical bolometric flare energies measured for the Bastille Day and Halloween 2003 events, while the 95\% and 99\% prediction intervals encompass the most extreme observed cases.
For the AR 12192 (October 2014), our estimated ribbon areas are also in good agreement with those measured by K2017 \cite{kazachenko_database_2017} from SDO/HMI and SDO/AIA observations.

For the historical events, where no direct magnetic or ribbon measurements exist, the empirically derived scaling provides the only means of estimating plausible flare energies.  
The Carrington event of 1859, with an inferred spot area of $\sim3100$~msh, corresponds to an expected energy of ${\sim}1\times10^{33}$~erg under the 95th-envelope relation and up to ${\sim}7\times10^{33}$~erg within the 99\% prediction interval.  
These values are broadly consistent with independent energy estimates based on geomagnetic and auroral effects, and comparable to the largest modern solar flares.  
The still larger 1947 sunspot group would, under the same framework, be capable of producing a flare of a few ${\times}10^{34}$~erg, approaching the lower end of the stellar ``superflare'' regime, 
consistent with solar–stellar comparisons \citep[e.g.,][]{Herbst_2021}.

Overall, the distribution of predicted energies suggests that:
\begin{itemize}
    \item The ``best'' relations reflect the energetics of regularly observed large solar flares ($10^{32}$--$10^{33}$~erg);
    \item The 95\% prediction intervals encompass the rarest but still physically plausible solar extremes (${\lesssim}10^{34}$~erg);
    \item The 99\% prediction intervals approach the statistical and physical upper limits, comparable to the most energetic events inferred from stellar and cosmogenic isotope evidence.
\end{itemize}

We therefore conclude that our empirical framework, calibrated on modern observations, provides a physically and statistically self-consistent extrapolation to the most extreme solar conditions.  
The Carrington and 1947 cases delineate the probable upper bound of solar flare energies, while the 99\% prediction interval marks the threshold beyond which only the very rarest ``superflare-like'' solar events may occur~--- if at all.
This strengthens the case that extreme solar particle events identified in cosmogenic-isotope records could indeed be driven by exceptionally rare but physically plausible solar flares and CMEs.

It is important to note that the quantitative estimates presented here carry systematic uncertainties.  
The method assumes that the largest historical sunspots had similar physical properties to those observed in the SDO era, such that the empirical scalings remain valid when extrapolated upward. 
Uncertainties arise at each step of the chain~--- from the scatter in the Chatzistergos et~al. \cite{chatzistergos_scrutinising_2022} spot-to-AR conversion, through the distribution of ribbon-to-AR area ratios, to the energy scaling itself.
We propagate these uncertainties conservatively, emphasising that our goal is not to provide tight error bars, but rather to establish plausible upper limits on solar flare energies.
Additional sources of uncertainty include projection and threshold effects in area measurements, temporal evolution of AR properties, as well as the intrinsic variability of flare productivity among active regions of similar size.  
The conversion between X-ray and bolometric flare energies, to which we compare our estimates, is also imperfectly constrained, and the derived bolometric energy can further depend on the flare location on the solar disc~--- events near the limb may appear weaker in total irradiance owing to partial occultation and anisotropic emission \cite{woods_contributions_2006}. 
While these factors may introduce discrepancies of up to two to three in absolute energy, they do not affect the relative consistency between the image-based and empirical approaches, nor the overall hierarchy of inferred flare energies across the sample.  
These caveats should therefore be kept in mind when interpreting the highest predicted values approaching the superflare regime.

\subsection{Role of nesting}
\label{sec:nesting}
So far, our analysis has treated each large active region (AR) in isolation. 
However, observations show that ARs on the Sun do not always emerge as spatially and temporally independent entities. 
Instead, they often appear in clusters, often referred to as ``nests of activity'' \cite{gaizauskas1983,brouwer1990,usoskin2005}, or sometimes also as ``active longitudes''. 
Such nests represent regions of preferred magnetic flux emergence, where new ARs repeatedly form close to pre-existing ones. 
Statistical studies suggest that about 40--60\% of sunspot groups are associated with such nests \cite{Pojoga_clustering_2002,Karapinar26}, and that the degree of nesting on the Sun increases with its activity level \cite{Karapinar26}.
For more active stars the degree of nesting appears to be even higher \cite{Isik2020_nesting}.

The relevance of nesting for extreme solar activity becomes clear when considering concrete cases. 
During the Halloween events of October--November 2003, two exceptionally large ARs emerged at nearly the same longitude, albeit in different hemispheres, see Fig.~\ref{fig_2003}. 
The larger of the two reached a sunspot area of $\sim 3500$~msh, while the second exceeded 2500~msh \cite{mandal_sunspot_2020}~--- that is larger than the July 2000 spot, see Tables~\ref{tab:events} and \ref{tab:full}. 
Although our analysis focuses on the larger AR, the contemporaneous presence of a second large region highlights the possibility that, had they emerged even closer together, their combined interaction could have created a much more magnetically complex environment than either region in isolation.

Recent events provide further illustration. 
The active region NOAA AR 13664 responsible for the May 2024 ``Mother’s Day'' (or ``Gannon'') event exhibited a complex evolution in which at least one additional AR, NOAA AR 13668, and potentially yet another smaller one emerged and merged into the pre-existing large region \cite{Dikpati25,kontogiannis2025}. 
This sequential merging, clearly visible in high-cadence magnetograms, enhanced the complexity and free-energy budget of the overall system, plausibly facilitating the occurrence of the extreme flare and CME associated with the event.
Yet another example of eruptions in a nested AR system (NOAA Active Region 11791 in July 2013) is discussed in \cite{Karpen_eruptions_2024}.

From a physical perspective, the clustering or merging of large ARs is likely to increase both the amount of magnetic flux available for reconnection \cite{Finley25} and the probability of large-scale, system-wide instabilities \cite{Toriumi2019}. 
In terms of our empirical framework, which links sunspot areas to flare energies via AR and ribbon areas, this implies that the ``single-AR'' scaling relations may represent conservative estimates. 
If two or more very large ARs interact or merge, the effective AR area and associated reconnection volume could exceed our single-region extrapolations, thereby allowing for flare energies beyond the limits inferred here.

While AR nesting offers a plausible alley to exceeding the limits implied by single-AR scaling relations, a quantitative treatment is not straightforward.
In particular, the effective ribbon area produced by nested or interacting active regions does not necessarily scale with the simple sum of their projected areas, as it depends sensitively on magnetic connectivity, reconnection geometry, and the temporal evolution of flux emergence.
As a heavily simplified illustration, if two large active regions were to reconnect in a fully coupled manner, the effective area participating in reconnection could approach that of the combined system, potentially shifting an event toward the upper tail of the energy-area relations shown in Fig.~\ref{fig_AR_energy}. 
A physically meaningful quantitative estimate of nesting therefore requires a dedicated treatment of magnetic topology and flux evolution, which is beyond the scope of the present empirical analysis.

In summary, nesting provides a natural pathway by which the Sun may occasionally produce flares and particle events stronger than predicted from the statistics of individual ARs. 
This highlights the importance of considering AR interactions when assessing the upper limits of solar eruptive activity.

\section{Summary and Conclusions}
\label{sec:conclude}

In this study, we have combined historical records of the largest sunspot groups with empirical relationships between sunspot areas, active-region (AR) areas, ribbon areas, and flare energies. 
Our goal was to provide constraints on the maximum possible flare energies that could arise from the most extreme sunspot groups in the historical record.
We emphasize that this analysis does not address the probability or expected frequency of such events.

Our approach is empirical and consists of a sequence of linked steps, each building on the previous one.  
We began by compiling records of the largest historical sunspot groups \cite{mandal_sunspot_2020}, which served as our starting point.  
Since sunspots represent only a fraction of the magnetic flux of an active region, we estimated the total AR area using two approaches: direct measurements from magnetograms and an empirical scaling relation \cite{chatzistergos_analysis_2020} that relates sunspot area to total AR area.  
Next, we used statistical relations between AR areas and flare ribbon areas, derived from the K2017 catalogue of SDO-era flares, to estimate the plausible ribbon areas for large historical ARs.  
Finally, we employed the empirical scaling \cite{kazachenko_database_2017} between ribbon areas and flare energies to infer the maximum plausible flare energy associated with each AR.

A key test case was AR\,12192 (October 2014), for which both K2017 catalogue data \cite{kazachenko_database_2017} and our independent AR estimates are available.  
We find that our AR and “best-estimate” predictions of ribbon areas agree well with the observed values.
The empirical prediction intervals then provide plausible upper bounds.  
This gives us confidence that the same methodology can be extended to the historical record.

We then examined several large sunspot groups from the M2020 \cite{mandal_sunspot_2020} catalogue, including the one which led to the October--November 2003 “Halloween” event and the great spot of April 1947, as well as the 1859 Carrington spot.  
For AR\,10486 in 2003, our estimates of the flare energy are broadly consistent with the bolometric energies inferred from TSI observations of the associated X25 flare \cite{woods_solar_2004}.
The agreement between modern validation cases and historical extrapolations lends confidence that the empirical framework can be used to constrain the realistic upper limits of solar flare energies, even in the absence of direct flare observations.
The 1947 spot, the largest in the last century, yields the highest possible flare energy within our framework, of the order of $10^{34}$~erg, the level typically considered for “superflare” scenarios.

Overall, we conclude that the largest historical sunspot groups, when scaled through modern empirical relations, are capable of powering flares with bolometric energies up to a few $\times 10^{34}$ erg.  
Such events lie at the very tail of the observed flare distribution but remain physically consistent with the properties of known active regions.  
We note, however, that nesting or clustering of multiple large ARs~--- as observed on both the Sun and Sun-like stars~--- may further enhance the likelihood and potential total energy output of extreme flares, beyond the limits inferred for individual ARs.

\ack{We thank the reviewers for their careful reading of the paper and helpful suggestions.
TC acknowledges funding from the European Research Council (ERC) under the European Union's Horizon 2020 research and innovation programme (grant agreement No. 101097844 — project WINSUN).
This study has made use of Smithsonian Astrophysical Observatory (SAO)/NASA's Astrophysics Data System (ADS; \url{https://ui.adsabs.harvard. edu/}) Bibliographic Services.
}

%%%%%%%%%% Insert bibliography here %%%%%%%%%%%%%%
%\bibliographystyle{RS}
%\bibliography{sample,Spots-and-Flares/full_lib_thes_cp}
%\begin{thebibliography}{9}
%\end{thebibliography}

\end{document}